\documentclass[12pt,a4paper]{article}
\usepackage{amsmath,amssymb}
\usepackage{graphicx}
\usepackage{color}

\begin{document}
\setlength{\baselineskip}{18pt}
\begin{titlepage}
\begin{flushright}
CTPU-17-32\\ 
EPHOU-17-012
\end{flushright}
\vskip .75in

\begin{center}
{\Large\bf 
Very low scale Coleman-Weinberg inflation with non-minimal coupling
}
\end{center}
\lineskip .75em
\vskip 1.5cm

\begin{center}
{\large Kunio Kaneta}$^1$, 
{\large Osamu Seto}$^{2,3}$,
{\large Ryo Takahashi}$^4$ \\

\vspace{1cm}

$^1${\it Center for Theoretical Physics of the Universe,\\Institute for Basic Science (IBS), Daejeon 34051, Korea}\\
$^2${\it Institute for International Collaboration,\\Hokkaido University, Sapporo 060-0815, Japan}\\
$^3${\it Department of Physics, Hokkaido University, Sapporo 060-0810, Japan}\\
$^4${\it Graduate School of Science, Tohoku University, Sendai 980-8578, Japan}\\

\vspace{10mm}
{\bf Abstract}\\[5mm]
{\parbox{13cm}{\hspace{5mm}
We study viable small-field Coleman-Weinberg (CW) inflation models with the help of non-minimal coupling to gravity. 
The simplest small-field CW inflation model (with a low-scale potential minimum) is incompatible with the cosmological constraint on the scalar spectral index. However, there are possibilities to make the model realistic. 
First, we revisit the CW inflation model supplemented with a linear potential term.
We next consider the CW inflation model with a logarithmic non-minimal coupling and illustrate that the model can open a new viable parameter space that includes the model with a linear potential term.
We also show parameter spaces where the Hubble scale during the inflation can be as small as $10^{-4} $ GeV, $1$ GeV, $10^4 $ GeV, and $10^8$ GeV
 for the number of $e$-folds of $40,~45,~50$, and $55$, respectively, with other cosmological constraints being satisfied.
}}
\end{center}
\end{titlepage}

\section{Introduction}
Inflation is one of the successful paradigms in modern cosmology that can address various cosmological issues~\cite{Starobinsky:1980te,Sato:1980yn,Guth:1980zm} and generate primordial perturbations~\cite{Mukhanov:1981xt,Hawking:1982cz,Starobinsky:1982ee,Guth:1982ec}.
The underlying particle physics is, however, still unclear, and it is indispensable in understanding physics in the early Universe.
To this end, in particular, it is legitimate to ask what is a consistent inflationary scenario for a specific physics model beyond the standard model of particle physics.

Two categories are often used to classify various inflationary models: large-field and small-field inflation, according to whether the inflaton field excursion during inflation exceeds the Planck scale or not.
Each class of models has its own virtues.
For instance, the large-field models have an advantage in the initial condition of inflation~\cite{Linde:1983gd,Brandenberger:1990wu}, whereas in the small-field models, inflation can take place with the inflaton field value well below the Planck scale, and hence, its field theoretical description is verified and well understood.

In small-field models with a symmetry-breaking-type potential, inflation takes place at the vicinity of the origin, and the inflaton field slowly rolls down toward the potential minimum located below the Planck scale.
From the normalization of temperature anisotropy of cosmic microwave background (CMB) radiation of $\mathcal{O}(10^{-5})$, the energy scale of small-field inflation models, which is equivalently the Hubble parameter during inflation, turns out to be rather small. Thus, a small-field inflation generally leads to a rather low reheating temperature.

Such a low-scale inflation and its resultant low reheating temperature are attractive from several viewpoints.
Here, we note several examples and those motivations.
First, in a Peccei-Quinn (PQ) extended model to solve the strong CP problem in the standard model of particle physics~\cite{Peccei:1977hh,Peccei:1977ur}, if the PQ symmetry is broken before or during inflation, axion fluctuations on the order of the Hubble parameter during inflation are generated and induce axion isocurvature perturbations~\cite{Axenides:1983hj,Seckel:1985tj,Linde:1985yf,Turner:1990uz}. 
To satisfy the stringent bound on axion isocurvature perturbation by the CMB temperature anisotropy, a small Hubble parameter during inflation, $\lesssim 10^7$ GeV, is required~\cite{Ade:2015lrj}.
Second, one of the most promising scenarios for generation of baryon asymmetry is the Affleck-Dine mechanism with a flat direction~\cite{Affleck:1984fy}.
An appropriate amount of baryon asymmetry can be generated by a flat direction lifted by a dimension-six operator for
a low reheating temperature $T_R$ of about $100$ GeV~\cite{Dine:1995kz}. Affleck-Dine baryogenesis by such flat directions is interesting because it provides a solution to the coincidence of energy densities between baryon and dark matter with the formation of $Q$-balls~\cite{Enqvist:1998en,Roszkowski:2006kw,Seto:2007ym}. 
Third, an issue of supersymmetric models in cosmology is the overproduction of the gravitino~\cite{Nanopoulos:1983up,Khlopov:1984pf,Ellis:1984eq}.
Because gravitino abundance produced through thermal scatterings is proportional to the reheating temperature after inflation, in order to avoid overproduction, the upper bound on the reheating temperature is imposed. For a recent estimation, see, e.g., Refs.~\cite{Kawasaki:2008qe,Cyburt:2009pg}.
Finally, the recently proposed relaxion mechanism~\cite{Graham:2015cka}, as a solution to the hierarchy problem of Higgs boson by utilizing a slowly rolling scalar field in the context of inflationary cosmology, also requires a very long period and a very low energy scale of inflation for a phase transition by QCD(-like) strong dynamics to take place during inflation, not only in the minimal model~\cite{Graham:2015cka,Kobayashi:2016bue,Choi:2016luu} but also in some extended models~\cite{Batell:2015fma,DiChiara:2015euo,Evans:2016htp,Evans:2017bjs}.
(However, for other extensions where the relaxion mechanism can work at relatively high scale, see e.g., Refs.~\cite{Hook:2016mqo,Higaki:2016cqb,Choi:2016kke}.)

In this paper, we pursue a possible realization of viable small-field inflationary models based on the Coleman-Weinberg (CW) model~\cite{Coleman:1973jx}.
In particular, we discuss how small inflation scale can be achieved in the CW model with some possible modifications.
The CW inflation model is a typical model of low-scale and small-field inflation~\cite{Linde:1981mu,Albrecht:1982wi,Shafi:1983bd}.
However, the original CW inflation model is doomed by the observed scalar spectral index~\cite{Barenboim:2013wra}, which is significantly larger than that of the model predictions.\footnote{
If the vacuum expectation value of the field is larger than the Planck scale, the CW potential might reproduce the consistent density perturbation~\cite{Okada:2014lxa} (for other attempts, see, e.g.~\cite{Shafi:2006cs,Rehman:2008qs,Kannike:2015kda,Marzola:2015xbh,Marzola:2016xgb,Artymowski:2016dlz}). However, this case belongs to the large-field model
 and is not discussed in this paper.}
Iso \textit{et al.} have proposed simple extensions to ameliorate this discrepancy~\cite{Iso:2014gka}.
In this paper, we revisit known examples of such extension, and explore further possibilities by considering other promising extensions.

The rest of this paper is organized as follows.
In Sec.~\ref{sec:CW inflation}, we first go over the models discussed in Ref.~\cite{Iso:2014gka} and move onto other possible extensions based on a non-minimal coupling to gravity.
We devote Sec.~\ref{sec:conclusion} to discussions and conclusions.

\section{Small-field Coleman-Weinberg inflation model}
\label{sec:CW inflation}
We study a class of small-field CW inflation where the inflaton starts to roll down from the vicinity of the origin to the potential minimum~\cite{Linde:1981mu,Albrecht:1982wi,Shafi:1983bd,Iso:2014gka}. The scalar potential for the inflaton $\phi$ is given by
\begin{eqnarray}
V(\phi)=\frac{A}{4}\phi^4\left(\ln\frac{\phi^2}{M^2}-\frac{1}{2}\right)+V_0,~~~
V_0=\frac{AM^4}{8},
\label{CW}
\end{eqnarray}
with a scale $M(<M_{\rm pl})$ and $M_{\rm pl}$ being the reduced Planck mass. $V_0$ is determined by the vanishing cosmological constant at the minimum. Derivatives of the potential with respect to $\phi$ are
\begin{eqnarray}
V'=A\phi^3\ln\frac{\phi^2}{M^2},~~~
V''=A\phi^2\left(2+3\ln\frac{\phi^2}{M^2}\right).
\end{eqnarray}
We find that the vacuum expectation value at the minimum is given by $\langle \phi\rangle =M$ and $V_0$ is obtained by $V(M)=0$ as shown above.
Thus, the slow roll parameters are calculated as
\begin{eqnarray}
\epsilon &=& \frac{M_{\rm pl}^2}{2}\left(\frac{V'}{V}\right)^2
         \simeq 32\left(\frac{M_{\rm pl}}{M}\right)^2\left(\frac{\phi}{M}\right)^6
                \left(\ln\frac{\phi^2}{M^2}\right)^2, \\
\eta &=& M_{\rm pl}^2\left(\frac{V''}{V}\right)
     \simeq 24\left(\frac{M_{\rm pl}}{M}\right)^2\left(\frac{\phi}{M}\right)^2
            \ln\frac{\phi^2}{M^2},
\end{eqnarray}
where $V\simeq V_0$ is utilized in a small-field region of $\phi$. The slow roll conditions can be satisfied when a field value of $\phi$ is small enough to satisfy $|\eta|<1$ for a given $M$. In this region, we find $\epsilon\ll|\eta|$ and $\phi\ll M$. This simple CW inflation model leads to a tiny tensor-to-scalar ratio ($r\simeq16\epsilon$), which is allowed by the current bound~\cite{Ade:2015lrj,Array:2015xqh} from cosmological observations. However, for $M\ll M_{\rm pl}$, the scalar spectral index ($n_s\simeq1+2\eta-6\epsilon$) from this model as $0.94\lesssim n_s\lesssim 0.95$ for $50\leq N_\ast\leq60$ does not fall into the allowed region for a tiny $r$ as   
\begin{eqnarray}
& 0.955 \lesssim n_s \lesssim 0.976 ~~~(68\% \mathrm{CL}),\label{nsbound1}  \\
& 0.949 \lesssim n_s \lesssim 0.982 ~~~(95\% \mathrm{CL}),
\label{nsbound2}
\end{eqnarray}
given by Planck TT+lowP data~\cite{Ade:2015lrj}. $N_\ast$ is the number of $e$-folds given by
\begin{eqnarray}
N_\ast=\frac{1}{M_{\rm pl}^2}\int_{\phi_{\rm end}}^{\phi_\ast}\frac{V}{V'}d\phi
 \simeq\frac{1}{M_{\rm pl}^2}\int_{\phi_{\rm end}}^{\phi_\ast}\frac{V_0}{A\phi^3\ln(\phi^2/M^2)}d\phi ,
\end{eqnarray}
with field value $\phi_\ast$ where the pivot scale $k_\ast$ exits from the Hubble radius. $\phi_{\rm end}$ denotes the field value at the end of inflation. Therefore, the low-scale CW inflation model must be modified to be consistent with cosmological observations.
In most studies, $N_*$ is taken to be about $50$ or $60$.
In fact, $N_*$  weakly depends on the energy scale of inflation and delay of reheating after inflation~\cite{Liddle:1993fq,Lyth:1998xn,Liddle:2003as} as
\begin{eqnarray}
N_* \simeq 62- \ln\frac{10^{16} \mathrm{ GeV}}{V_*^{1/4}} - \frac{1}{3}\ln\frac{V_*^{1/4}}{\rho_R^{1/4}} ,
\end{eqnarray}
with the energy density at the reheating $\rho_R$ and the energy density at the moment of the pivot scale horizon crossing during inflation $V_*$ for the standard thermal history after inflation in which the Universe becomes the matter dominated with the equation of state $w=0$ during the coherent oscillation of inflaton after inflation, followed by the radiation-dominated Universe. Here, we used $V_\ast \simeq V(\phi_{\rm end})$.
Now, $N_*$ is a function of $V_*$ and $\rho_R$. In the following analysis, because we are interested in very low scale CW inflation, we vary $N_*$ from $40$ to $55$ under the condition $\rho_R \leq V_*$.

\subsection{Fermion condensates}

A possibility to increase $n_s$ is the introduction of a linear term, which can  be generated by a fermion condensation in the inflaton potential discussed in Ref.~\cite{Iso:2014gka}. In the work, two examples that induce a linear term have been shown. One is the condensation of right-handed neutrinos $N$, which couples to $\phi$ through a Yukawa interaction $y_N\phi\bar{N}^cN$. The other one is the chiral condensation, which generates a linear term as $C_hh$ in the Higgs ($h$) potential. Then, the mixing between the Higgs and inflaton induces a linear term in the inflaton potential. In both cases, a linear term, $C\phi$, in the inflaton potential can be induced from a fermion condensate.

The potential (\ref{CW}) is changed to 
\begin{eqnarray}
V(\phi)=\frac{A}{4}\phi^4\left(\ln\frac{\phi^2}{M^2}-\frac{1}{2}\right)-C\phi
        +V_0,~~~
V_0=\frac{AM^4}{8} .
\label{CWl}
\end{eqnarray}
$V'$ and $\epsilon$ are also modified to 
\begin{eqnarray}
V'=A\phi^3\ln\frac{\phi^2}{M^2}-C,~~~
\epsilon\simeq32\left(\frac{M_{\rm pl}}{M}\right)^2
                \left[\left(\frac{\phi}{M}\right)^3
                      \ln\frac{\phi^2}{M^2}-\frac{C}{AM^3}\right]^2.
\end{eqnarray}
$V''$ is unchanged, but $N_\ast$ is modified as
\begin{eqnarray}
N_\ast\simeq\frac{1}{M_{\rm pl}^2}\int_{\phi_{\rm end}}^{\phi_\ast}\frac{V_0}{A\phi^3\ln(\phi^2/M^2)-C}d\phi.
\label{Nast}
\end{eqnarray}
Thus, the relation between $n_s$ and $N_\ast$ changes from the original CW inflation. 

\begin{figure}
\begin{center}
\includegraphics[scale=.8]{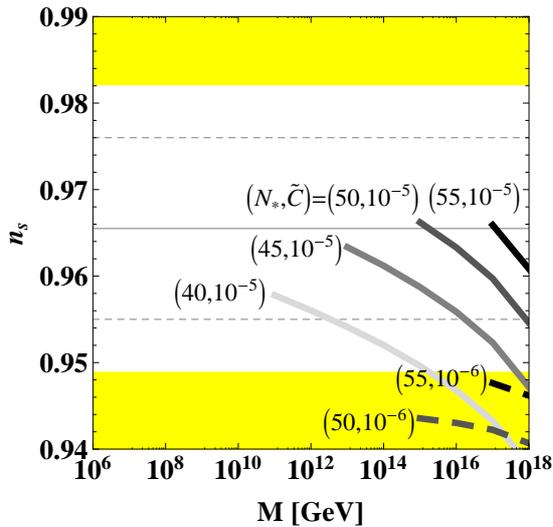}
\end{center}
\caption{Value of $n_s$ in the CW model with fermion condensates.
Black, dark gray, gray, and light gray curves correspond to $N_\ast=55,~50,~45$, and $40$ cases, respectively. Solid and dashed curves are $\tilde{C}=10^{-5}$ and $10^{-6}$, respectively, where $\tilde{C}$ is a dimensionless parameter defined as $\tilde{C}\equiv C(M_{\rm pl}/M)^3/(AM^3)$. The center value and $68\% \mathrm{CL}$ and $95\% \mathrm{CL}$ range of $n_s$ are shown by solid and dashed lines and shaded region, respectively.}
\label{fig1}
\end{figure}

For all figures in this paper, we normalize the amplitude of curvature perturbation at the pivot scale as $A_s=2.196\times10^{-9}$. 
Fig.~\ref{fig1} shows $n_s$ as a function of $M$ in the model with fermion condensates.
In the figure, black, dark gray, gray, and light gray curves correspond to $N_\ast=55, 50, 45$ and $40$ cases, respectively. Solid and dashed curves are cases of $\tilde{C}=10^{-5}$ and $10^{-6}$, respectively, where $\tilde{C}$ is a dimensionless parameter defined as $\tilde{C}\equiv C(M_{\rm pl}/M)^3/(AM^3)$. 
The end of each curve corresponds to the case of the instantaneous reheating, $\rho_R = V_*$, which gives the maximal reheating temperature after inflation.
The horizontal solid line denotes the center value of $n_s$ from the cosmological observation~\cite{Ade:2015lrj} as $n_s=0.9655$; its $1 \sigma$ range~(\ref{nsbound1}) is indicated by dashed lines, and shaded regions are outside of the bound~(\ref{nsbound2}). 
We find that the model has viable parameter space consistent with cosmological observations for
 $\tilde{C} = \mathcal{O}(10^{-5})$, with $40\lesssim N_\ast\lesssim 55$ in the broad region of $M \gtrsim 10^{11}$ GeV. 
The additional contribution of $C$ in the denominator of Eq. (\ref{Nast}) changes the value of $N_\ast$, but it does not change the magnitude of $\epsilon$, and thus the tensor-to-scalar ratio remains tiny as in the case of the original CW inflation model.

\begin{figure}
\begin{center}
\includegraphics[scale=0.8]{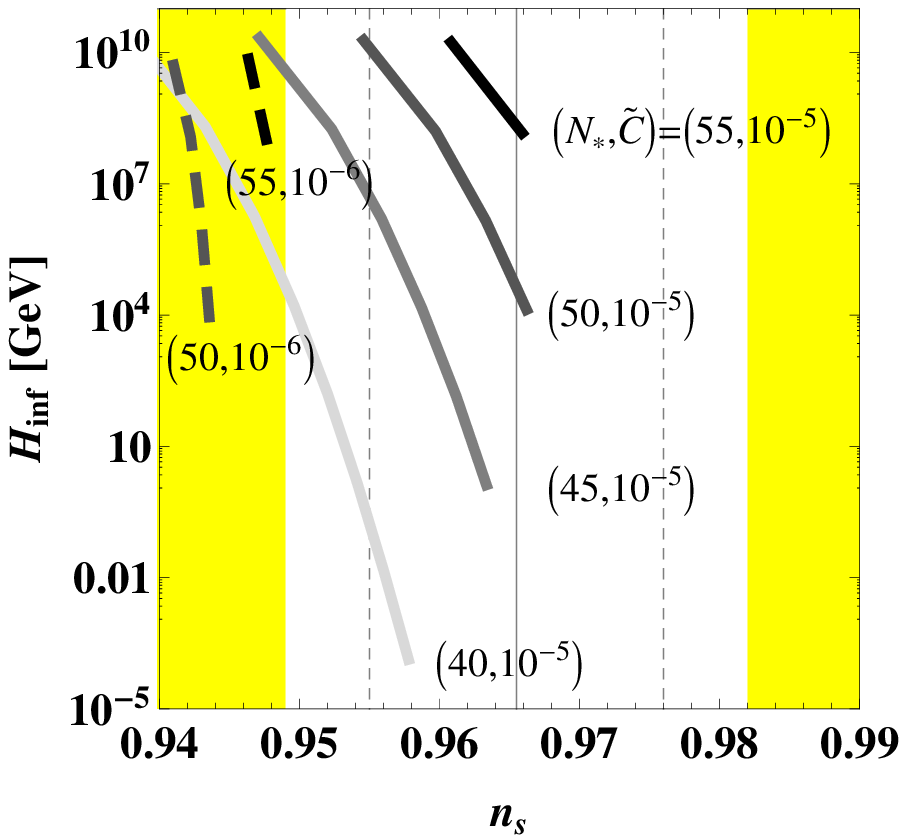}\hspace{1cm}
\includegraphics[scale=0.78]{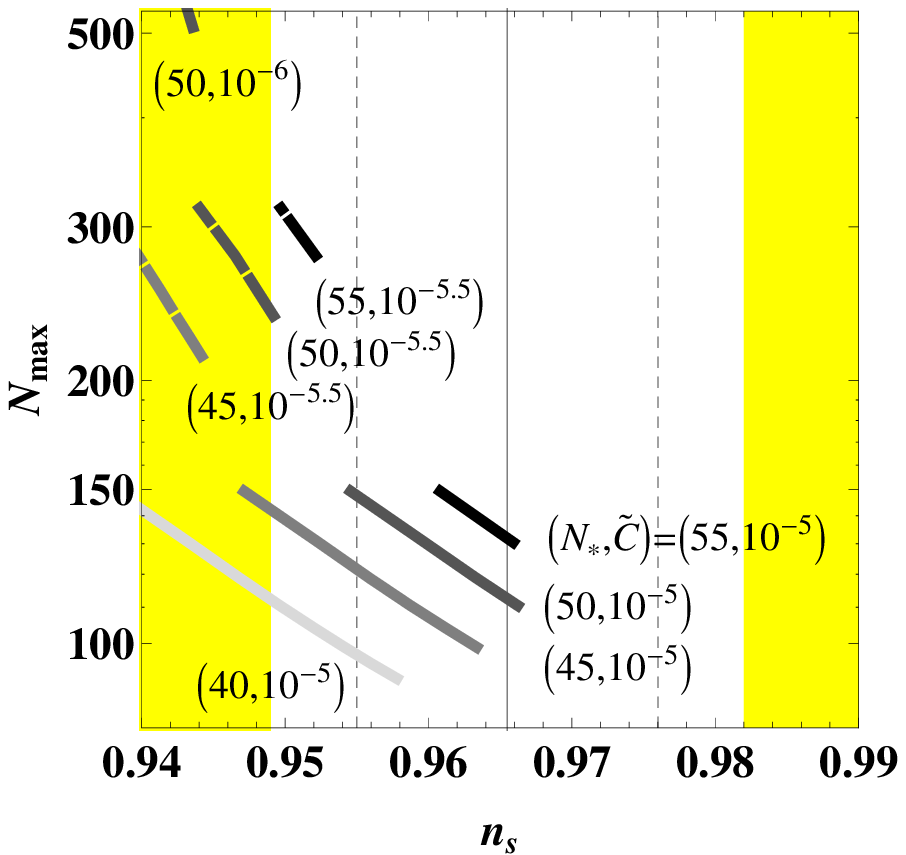}
\end{center}
\caption{Inflation scale (left) and maximal value of the number of $e$-folds (right) in this model. Solid, long dashed, and dashed curves on the right are $\tilde{C}=10^{-5}$, $10^{-5.5}$, and $10^{-6}$, respectively. (On the right figure, a dashed curve only appears in the upper-left corner.) The other meanings of curves and region in the figure are the same as those in Fig.~\ref{fig1}.} 
\label{fig2}
\end{figure}

Next, we discuss the inflation scale, that is, the Hubble scale during the inflation,  and maximal value of the number of $e$-folds during slow roll phase $N_{\rm max}$. The Hubble scale during the inflation $H_{\rm inf}$ is approximated as $H_{\rm inf}\simeq\sqrt{V_0/3M_{\rm pl}^2}$ in the model. $N_{\rm max}$ could be an interesting quantity from the viewpoint of the relaxion scenario as stated in the Introduction. $N_{\rm max}$ is defined as
\begin{eqnarray}
N_{\rm max}\equiv\int_{\phi_{\rm end}}^{\phi=0}\frac{V}{V'}d\phi,
\end{eqnarray}
in this class of small-field inflation model. The values in this model are shown in Fig.~\ref{fig2}.
Solid, long dashed and dashed curves in the right panel of Fig.~\ref{fig2} are $\tilde{C}=10^{-5}, ~10^{-5.5}$ and $10^{-6}$, respectively. Curves and regions in the figure are the same as those in Fig.~\ref{fig1}. We find that possible inflation scale is $H_{\rm inf}\gtrsim 10^{-4}$ GeV, $1$ GeV, $10^4$ GeV, and $10^8$ GeV for $N_\ast=40,~45, ~50$ and $55$ for $\tilde{C}=10^{-5}$, respectively. $M$ as large as $10^{18}$ GeV is available for $A\sim\mathcal{O}(10^{-14})$, and $H_{\rm inf}$ can be reduced to be as low as $10^{-4}$ GeV at $M=10^{11}$ GeV. Interestingly, this type of model can also realize small inflation scale compared with the usual large-field inflation models. The maximal value of the number of $e$-folds is $N_{\rm max}\simeq110,~140,~240$, and $310$ for $N_\ast=40,~45,~50$, and $55$, respectively, with an appropriate value of $\tilde{C}$.
Thus, enormous number of $e$-folds, which may be preferred in some relaxion models, cannot be realized in the model due to the absence of an extremely flat region such as a stationary point.

\subsection{Non-minimal coupling to gravity}

Let us now discuss another possible realization of a viable small-scale CW inflation, where we introduce a non-minimal coupling of the inflaton to gravity.
In Ref.~\cite{Iso:2014gka}, a non-minimal coupling to gravity of $\mathcal{L_\xi}=-\xi\bar\phi^2\mathcal{R}/2$ with $\mathcal{R}$ being the Ricci scalar and $\bar\phi$ being the Jordan frame inflaton field, has been discussed, but it was concluded that this term cannot make the original CW inflation viable.\footnote{Furthermore, we have also checked that introductions of cubic ($\xi\bar\phi^3\mathcal{R}/M$) and quartic ($\xi\bar\phi^4\mathcal{R}/M^2$) terms in the original CW inflaton potential do not work. As $\bar\phi$ is taken as small values during the inflation, higher terms than the quadratic one do not drastically change the properties of the original CW inflation model.} 

Instead of utilizing the quadratic coupling to gravity, we introduce a logarithmic term of non-minimal coupling to gravity.
Such a form of the non-minimal coupling may be obtained by incorporating quantum corrections to the $\bar\phi^2\mathcal{R}$ term~\cite{DeSimone:2008ei}{\footnote{See also~\cite{Buchbinder:1986yh,Elizalde:1993ee,Salvio:2014soa,Kannike:2015apa,Salvio:2017xul,Karam:2017zno}}}, and here we parametrize it as 
\begin{eqnarray}
	\mathcal{L}=-\xi M\bar{\phi}\mathcal{R}\left(\ln\frac{\bar{\phi}}{M}-c\right) ,
\label{non-minimal}
\end{eqnarray}
in the Jordan frame. In this case, the potential (\ref{CW}) is changed to 
\begin{eqnarray}
V(\bar{\phi})\rightarrow 
V_E(\bar{\phi}(\phi))=\frac{V(\bar{\phi}(\phi))}{\Omega^4(\bar{\phi}(\phi))}
\simeq \frac{A}{4}\bar{\phi}^4
       \left(\ln\frac{\bar{\phi}^2}{M^2}-\frac{1}{2}\right)
       +V_0\left(1+\frac{2\xi M\bar{\phi}}{M_{\rm pl}^2}
                   \left(\ln\frac{\bar{\phi}}{M}-c\right)\right), 
\label{Vlog2}
\end{eqnarray}
in the Einstein frame where $\Omega^2\equiv1-\frac{\xi M\bar{\phi}}{M_{\rm pl}^2}\ln(\bar{\phi}/M-c)$, and $\phi$ is a canonically normalized inflaton field in this frame. Although one can safely approximate $\Omega^2\simeq1$ and $d\phi/d\bar{\phi}\simeq1$ with $V(\bar{\phi})\simeq V_E(\bar{\phi}(\phi))\simeq V_0$ in a small-field region, the additional term induced from the logarithmic form of non-minimal coupling to gravity cannot be negligible for $V_E'$ and $V_E''$ in a certain parameter space. In particular, the model with a larger $c$ gives similar predictions from the CW model with a linear term discussed in the previous subsection. 

Taking derivatives with respect to $\phi$, we have
\begin{eqnarray}
&&V_E'(\bar{\phi}(\phi))
  =\frac{dV_E}{d\bar{\phi}}\frac{1}{\frac{d\phi}{d\bar{\phi}}}
  \simeq A\bar{\phi}^3\ln\frac{\bar{\phi}^2}{M^2}+\frac{A\xi M^5}{4M_{\rm pl}^2}
         \left(\ln\frac{\bar{\phi}}{M}-c+1\right),  \\
&&V_E''(\bar{\phi}(\phi))
  \simeq A\bar{\phi}^2\left(2+3\ln\frac{\bar{\phi}^2}{M^2}\right)
         +\frac{A\xi M^5}{4M_{\rm pl}^2\bar{\phi}}.
\end{eqnarray}
Then, the slow roll parameters are calculated as
\begin{eqnarray}
\epsilon &=& \frac{M_{\rm pl}^2}{2}\left(\frac{V_E'}{V_E}\right)^2
         \simeq 32\left(\frac{M_{\rm pl}}{M}\right)^2
                \left[\left(\frac{\phi}{M}\right)^3\ln\frac{\phi^2}{M^2}
                +\frac{\xi}{4}\left(\frac{M}{M_{\rm pl}}\right)^2
                 \left(\ln\frac{\phi}{M}-c+1\right)\right]^2, \\
\eta &=& M_{\rm pl}^2\left(\frac{V_E''}{V_E}\right)
     \simeq 24\left(\frac{M_{\rm pl}}{M}\right)^2
            \left[\left(\frac{\phi}{M}\right)^2\ln\frac{\phi^2}{M^2}
            +\frac{\xi}{12}
            \left(\frac{M}{M_{\rm pl}}\right)^2\left(\frac{M}{\phi}\right)\right],
\end{eqnarray}
where $V_E\simeq V_0$ is utilized in the small-field region. $N_\ast$ becomes
\begin{eqnarray}
N_\ast\simeq\frac{1}{M_{\rm pl}^2}\int_{\phi_{\rm end}}^{\phi_\ast}
           \frac{V_0}
                {\left[
                 A\bar{\phi}^3\ln(\bar{\phi}^2/M^2)+\frac{A\xi M^5}{4M_{\rm pl}^2}
                 \left(\ln\frac{\bar{\phi}}{M}-c+1\right)\right]\frac{d\phi}{d\bar{\phi}}}d\phi,
\label{Nast2}
\end{eqnarray}
whereas $N_{\rm max}$ is also changed to
\begin{eqnarray}
N_{\rm max}\equiv\int_{\phi_{\rm end}}^{\phi=0}\frac{V_E}{V_E'}d\phi.
\end{eqnarray}

\begin{figure}
\begin{center}
\includegraphics[scale=0.9]{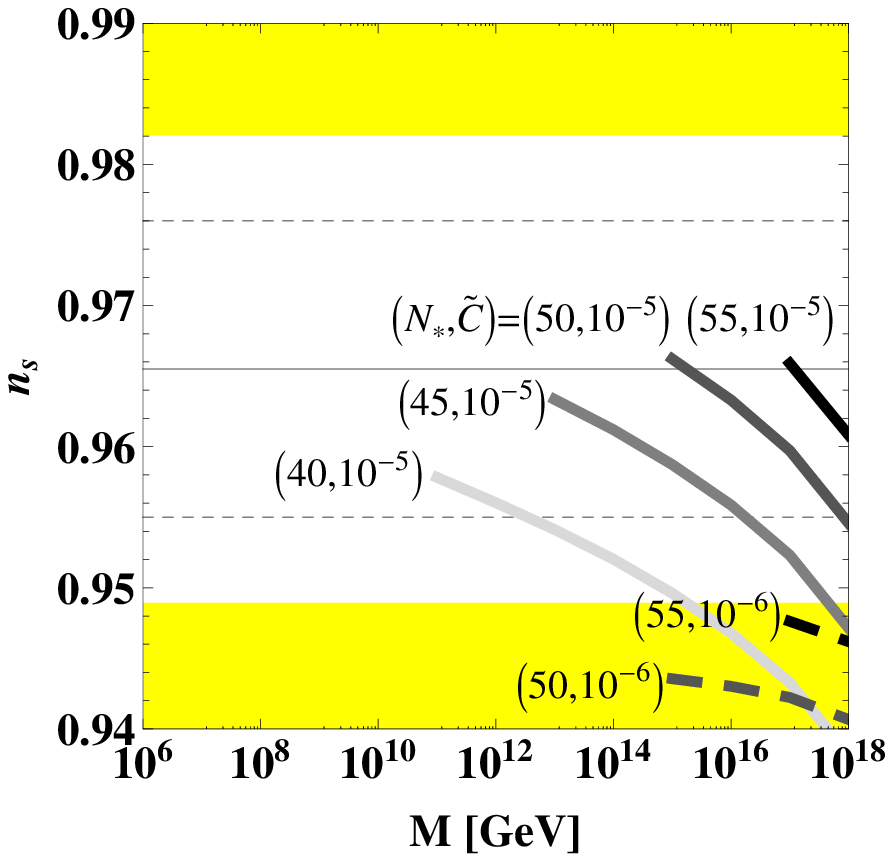}
\includegraphics[scale=0.9]{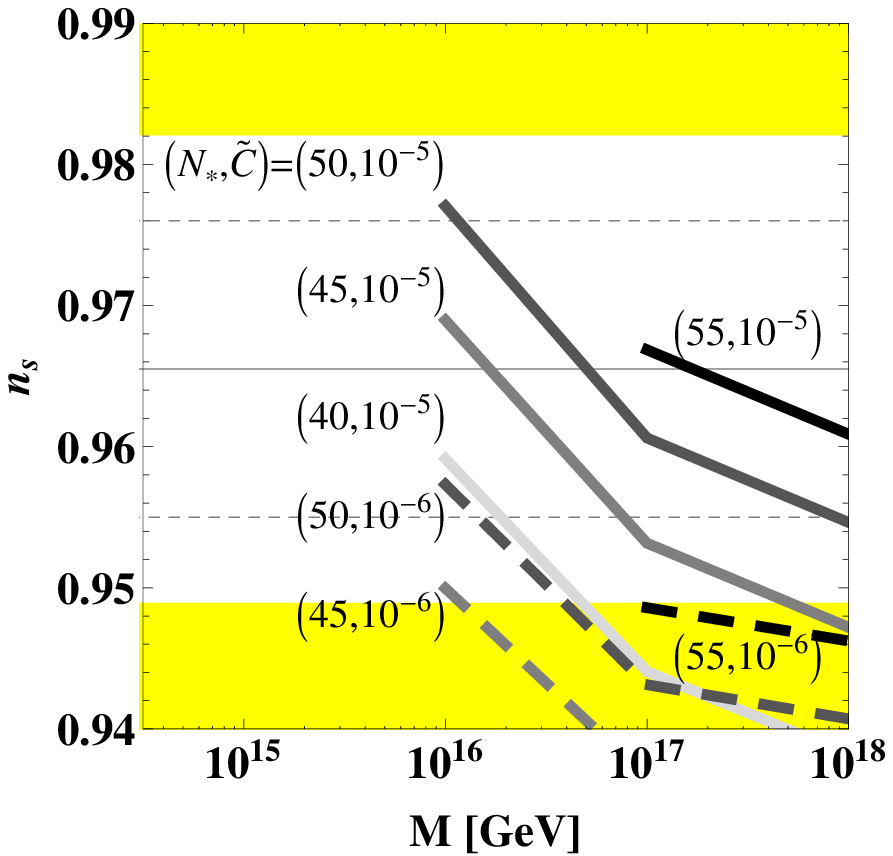}
\end{center}
\caption{Value of $n_s$ in the CW model with a logarithmic form of non-minimal coupling to gravity. Left and right panels correspond to $\xi=10^{-16}$ and $10^{-8}$, respectively. The meanings of curves and region in the figures are the same as those in Fig.~\ref{fig1}.} 
\label{fig3}
\end{figure}

The resultant $n_s$ in the model with a logarithmic form of non-minimal coupling to gravity is shown in Fig.~\ref{fig3}. 
Left and right panels correspond to $\xi=10^{-16}$ and $10^{-8}$, respectively. Lines and regions in the figures represent the same meaning as those in Fig.~\ref{fig1}. 
Similar to the case of the CW model with a linear term, we define 
\begin{eqnarray}
&& \tilde{C} \equiv \left(\frac{M_{\rm pl}}{M}\right)^3 \frac{C}{AM^3},  \\
&& C \equiv \frac{cA\xi M^5}{4 M_{\rm pl}^2}, \label{C}
\end{eqnarray}
so that this parametrization leads to the $-C\phi$ term in Eq.~(\ref{Vlog2}). One can approximate Eq.~(\ref{Vlog2}) as
\begin{eqnarray}
V_E(\bar{\phi}(\phi))\simeq
\left\{
\begin{array}{lll}
\frac{A}{4}\bar{\phi}^4\left(\ln\frac{\bar{\phi}^2}{M^2}-\frac{1}{2}\right)-C\phi+V_0 & \mbox{ for } & |\ln(\bar{\phi}/M)|\ll c \\
\frac{A}{4}\bar{\phi}^4\left(\ln\frac{\bar{\phi}^2}{M^2}-\frac{1}{2}\right)+\frac{C\bar{\phi}}{c}\ln\frac{\bar{\phi}}{M}+V_0 & \mbox{ for } & |\ln(\bar{\phi}/M)|\gg c\\
\end{array}
\right.,
\label{app}
\end{eqnarray}
with the dimension-full parameter $C$ defined in Eq.~(\ref{C}). The condition $|\ln(\bar{\phi}/M)|\ll c$ corresponds to
\begin{eqnarray}
\mathcal{O}(10^{1-2})\times\frac{\xi M_{\rm pl}}{4\tilde{C}}\ll M,
\end{eqnarray}
with the use of the parameter $\tilde{C}$ where a coefficient $\mathcal{O}(10^{1-2})$ corresponds to $|\ln(\bar{\phi}/M)|$. In this region, the model gives a similar prediction to the CW model with a linear term. On the other hand, the term $(2\xi MV_0\bar{\phi}/M_{\rm pl}^2)\ln(\bar{\phi}/M)$ in Eq.~(\ref{Vlog2}), which is the $(C\bar{\phi}/c)\ln(\bar{\phi}/M)$ term in Eq. (\ref{app}), becomes effective compared to the linear term in the region of $M\ll\mathcal{O}(10^{1-2})\times(\xi M_{\rm pl})/(4\tilde{C})$. Such regions appear around $M\lesssim10^{17}$ GeV in the right panel of Fig.~\ref{fig3}; that is, the logarithmic term of Eq.~(\ref{non-minimal}) becomes dominant at $M\lesssim10^{17}$ GeV. We find that the model can have parameter space where constraints from cosmological observations can be satisfied when one takes $10^{-6}\lesssim\tilde{C}\lesssim10^{-5}$ and $\xi\lesssim10^{-8}$ for $40\leq N_\ast \leq 55$ in the broad region of $M$. Interestingly, the effect of logarithmic contribution in non-minimal coupling can give larger $n_s$ in the region of $45\lesssim N_\ast\lesssim55$ compared to the model of fermion condensates. The additional contribution of $\xi$ in the denominator of Eq. (\ref{Nast2}) changes the value of $N_\ast$ but it does not change the magnitude of $\epsilon$, so the tensor-to-scalar ratio remains tiny.

\begin{figure}
\begin{center}
\includegraphics[scale=0.9]{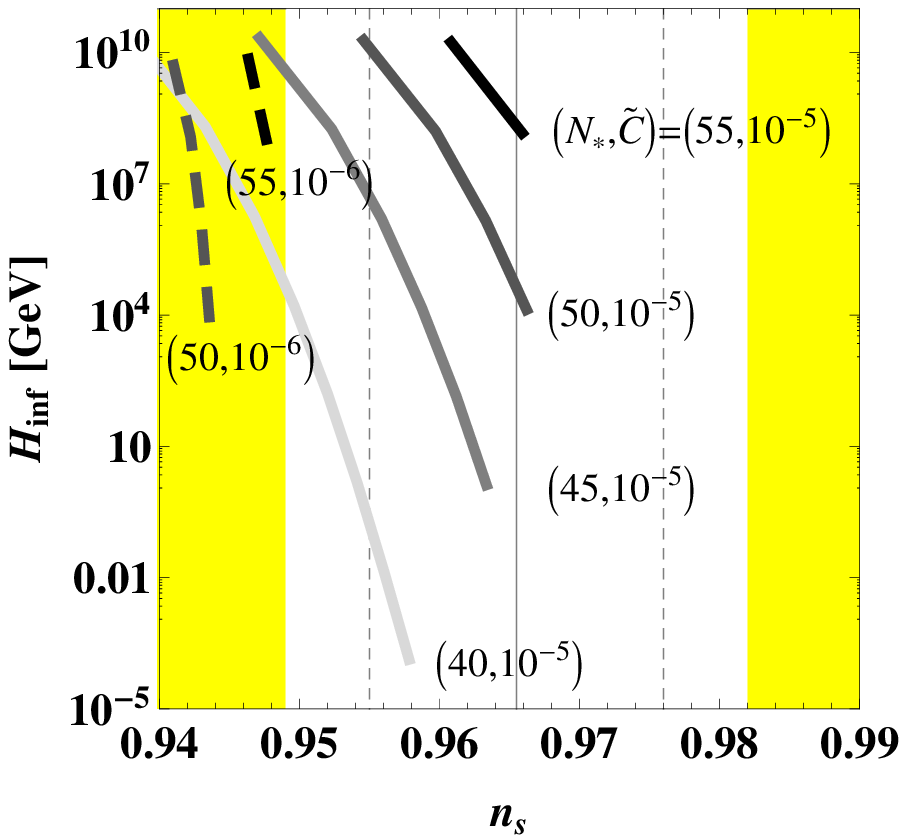}
\includegraphics[scale=0.9]{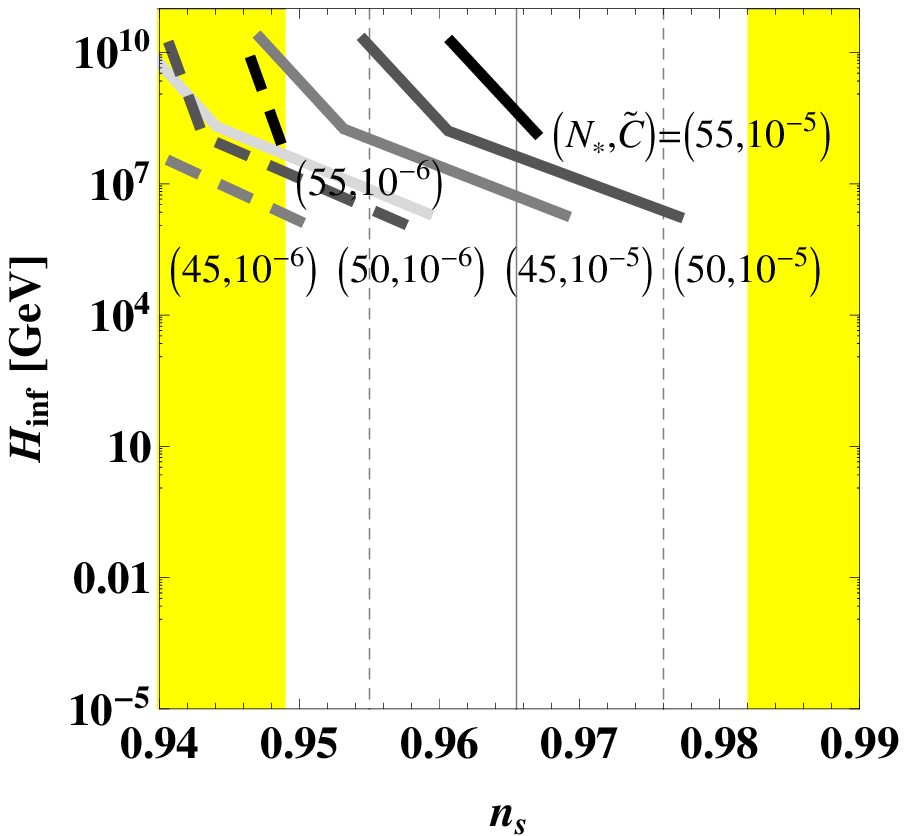}
\end{center}
\caption{$H_{\rm inf}$ in the CW model with a logarithmic form of non-minimal coupling to gravity. Left and right panels correspond to $\xi=10^{-16}$ and $10^{-8}$, respectively. The meanings of curves and region in the figures are the same as those in Fig.~\ref{fig1}.} 
\label{fig4}
\end{figure}

Figure \ref{fig4} shows $H_{\rm inf}$ as a function of $n_s$ for fixed $\xi,~N_*$, and $\tilde C$.
The curves and region in the figures are the same as those in Figs.~\ref{fig1} and \ref{fig3}. We find, in the left panel, that the possible inflation scale, $H_{\rm inf}\gtrsim10^{-4}$ GeV, $1$ GeV, $10^4$ GeV, and $10^8$ GeV for $N_\ast=40,~45,~50$, and $55$ with an appropriate value of $\tilde{C}$ for a smaller $\xi$, is the same as that in the CW model with a linear term. For a larger $\xi$ shown in the right panel, we find that the change of the curve around $H_{\rm inf} = 10^8$ GeV, where the logarithmic term becomes important, and increases $n_s$ more significantly than the linear term. 

\begin{figure}
\begin{center}
\includegraphics[scale=0.9]{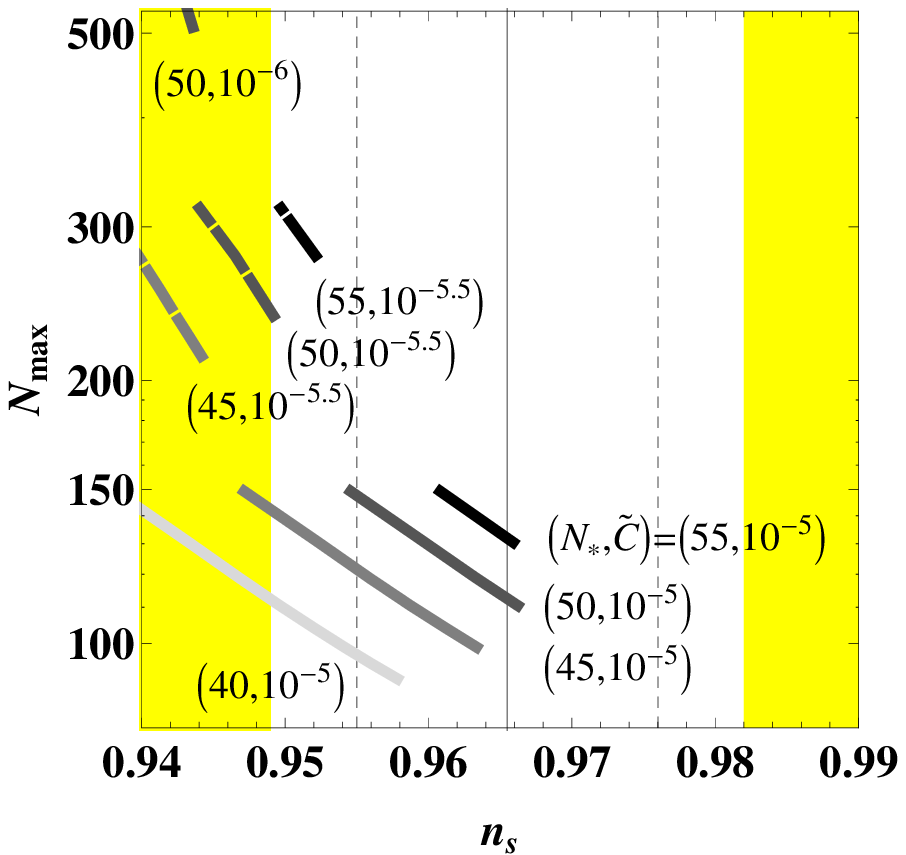}
\includegraphics[scale=0.9]{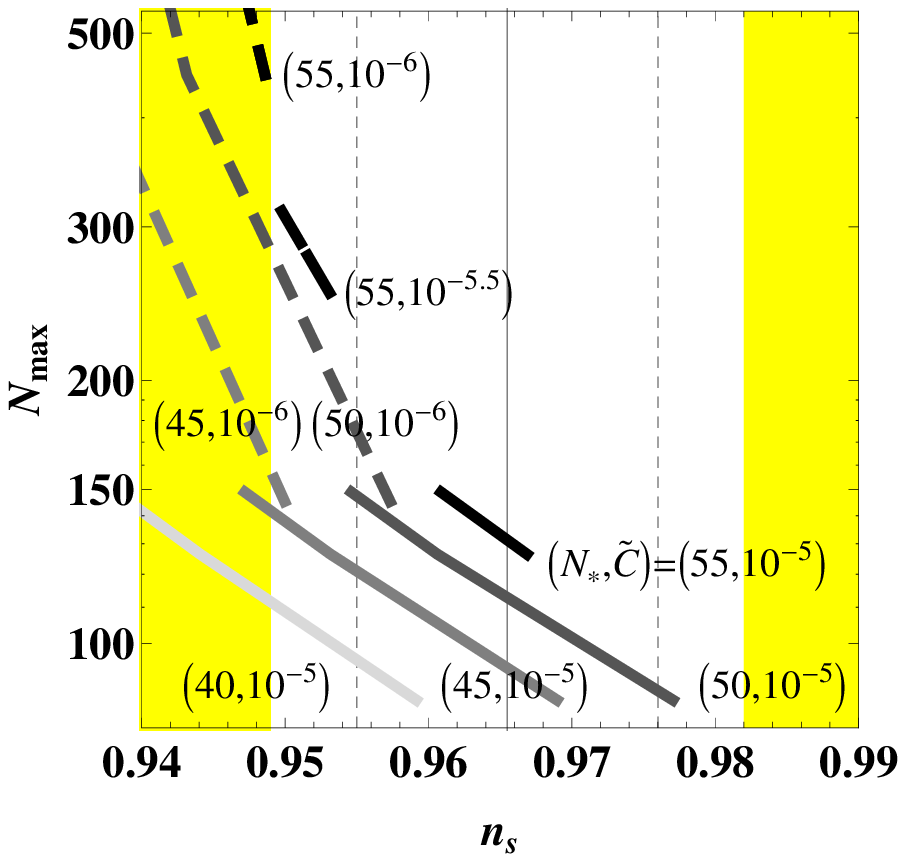}
\end{center}
\caption{$N_{\rm max}$ in the CW model with a logarithmic form of non-minimal coupling to gravity. Left and right panels correspond to $\xi=10^{-16}$ and $10^{-8}$, respectively. The meanings of curves and region in the figures are the same as those in the right panel of Fig.~\ref{fig2}.} 
\label{fig5}
\end{figure}

We also evaluate $N_{\rm max}$ as shown in Fig.~\ref{fig5}.
The maximal value of the number of $e$-folds is $N_{\rm max}\simeq 110,~140,~300$, and $310$ for $N_\ast=40,~45,~50$, and $55$, respectively, with an appropriate value of $\tilde{C}$. The maximal value of $N_\ast$ is similar to that in the case of the CW model with a linear term.

\section{Conclusions}
\label{sec:conclusion}

The CW potential for an inflaton realizes a small-field inflation but the current bound on $n_s$ from cosmological observation rules out the simplest small-field CW inflation model with smaller $M$. Thus, some modifications are necessary for such models to be consistent with cosmological observations.

An introduction of a linear term in the inflaton potential, which can be induced from fermion condensate, has been proposed to make the model realistic. In this work, first we have revisited this model with a linear term. In particular, we have investigated the inflation scale, that is the Hubble scale during inflation $H_{\rm inf}$, and maximal value of number of $e$-folds $N_{\rm max}$ in parameter space where cosmological bounds are satisfied. These two quantities would be relevant to various cosmological scenarios or problems. 
The lowest possible lowest inflation scale is $H_{\rm inf}\gtrsim 10^{-4}$ GeV, $1$ GeV, $10^4$ GeV and $10^8$ GeV for $N_\ast=40,~45,~50$, and $55$, respectively, with $\tilde{C} = 10^{-5}$, and the maximal value of the number of $e$-folds is $N_{\rm max}\simeq110,~140,~240$, and $310$ for $N_\ast=40,~45,~50$, and $55$, respectively, with an appropriate value of $\tilde{C}$.

Next, we have proposed another possible realization of the small-scale CW inflation, where the linear and logarithmic term of non-minimal coupling is introduced. This type of model also includes a parameter space where the model becomes similar to the CW model with a linear term. Regarding a possible inflation scale, we find $H_{\rm inf}\gtrsim 10^{-4}$ GeV, $1$ GeV, $10^4$ GeV, and $10^8$ GeV for $N_\ast=40,~45,~50$, and $55$, respectively, for $\tilde{C} = 10^{-5}$ with small $\xi$. Thus, one can realize small $H_{\rm inf}$ such as $10^{-4}$ GeV at $M=10^8$ GeV. Lower bounds depend on the magnitude of $\xi$ as $H_{\rm inf}\gtrsim 10^{-4}~(10^6)$ GeV for larger $\xi$ as $\xi=10^{-16}~(10^{-8})$. The model also gives the maximal value of the number of $e$-folds as $N_{\rm max}\simeq110,~140,~300$, and $310$ for $N_\ast=40,~45,~50$, and $55$, respectively. The possible maximal value of $N_\ast$ is similar that in the case of the CW model with a linear term.

In summary, the logarithmic non-minimal coupling can help make the small-scale CW inflation viable by increasing $n_s$. The non-minimal coupling can also realize a small inflation scale. In addition, motivated by the relaxion scenario, we have estimated the maximal number of $e$-folds, $N_{\rm max}$, which turns out to be $\mathcal{O}(100)$ and cannot be so enormous as required in relaxion models. The summary of possible additional terms to make the original CW model realistic is given in Tab.~\ref{tab1}.
\begin{table}
\begin{center}
\begin{tabular}{l||c|c|c|c}
\hline
Additional term \textbackslash~$N_\ast$ & 40 & 45 & 50 & 55 \\
\hline\hline
log & ($10^{-4}$ GeV, 110) & (1 GeV, 140) & ($10^4$ GeV, 300) & ($10^8$ GeV, 310) \\
\hline
linear & ($10^{-4}$ GeV, 110) & (1 GeV, 140) & ($10^4$ GeV, 240) & ($10^8$ GeV, 310) \\
\hline 
\end{tabular}
\end{center}
\caption{Summary of minimal values of $H_{\rm inf}$ ($\equiv H_{\rm inf}^{\rm min}$) and the number of $e$-folds, ($H_{\rm inf}^{\rm min}$, $N_{\rm max}$), for possible additional terms and $N_\ast$.}
\label{tab1}
\end{table}

\section*{Acknowledgments}

This work was supported by IBS under Project Code IBS-R018-D1.


\end{document}